\documentclass[10pt,twocolumn]{article} 

\usepackage[utf8]{inputenc} 

\usepackage{geometry} 
\usepackage{graphicx} 

\usepackage{cleveref}
\usepackage{chngpage}
\usepackage{booktabs} 
\usepackage{array} 
\usepackage{paralist} 
\usepackage{verbatim} 
\usepackage{subfig} 
\usepackage{amsmath}
\usepackage{mathtools} 
\usepackage{xcolor}
\usepackage{textcomp}
\usepackage{amsfonts}
\usepackage{float}
\usepackage{epigraph}
\usepackage{amsthm}
\usepackage{blindtext}
\usepackage{breqn}
\usepackage{marginnote}
\usepackage[toc,page]{appendix}
\usepackage{appendix}

\theoremstyle{definition}

\theoremstyle{remark}

\usepackage{comment}
\usepackage{amssymb}
\usepackage{ifthen}
\usepackage{color}
\specialcomment{notes}{\begingroup\sffamily\color{brown}}{\endgroup}

\usepackage[english]{babel}

\usepackage{fancyhdr} 
\usepackage{listings}
\lstdefinelanguage{IDL}{}
\lstdefinelanguage{JavaScript}{
  keywords={typeof, new, true, false, catch, function, return, null, catch,
    switch, var, if, in, while, do, else, case, break, eval, continue},
  keywordstyle={\ttfamily\color{dkviolet}},
  ndkeywords={class, export, boolean, throw, implements, import, this},
  ndkeywordstyle={\ttfamily\color{dkgreen}},
  identifierstyle={\ttfamily\color{black}},
  sensitive=true,
  comment=[l]{//},
  morecomment=[s]{/*}{*/},
  commentstyle={\ttfamily\color{dkgreen}},
  stringstyle=\ttfamily,
  morestring=[b]',
  morestring=[b]",
  columns=[l]fullflexible,
  extendedchars=true,
  basicstyle=\small,
  showstringspaces=false,
  showspaces=false,
  tabsize=2,
  breaklines=true,
  showtabs=false,
  captionpos=b
}[keywords,comments,strings]

\lstnewenvironment{jsl}{\lstset{language=JavaScript,basicstyle=\normalsize}}{}
\lstnewenvironment{jsls}{\lstset{language=JavaScript,basicstyle=\scriptsize}}{}
\lstnewenvironment{jsll}{\lstset{language=JavaScript,basicstyle=\small}}{}

\definecolor{ltblue}{rgb}{0,0.4,0.4}
\definecolor{dkblue}{rgb}{0,0.1,0.6}
\definecolor{dkgreen}{rgb}{0,0.4,0}
\definecolor{dkviolet}{rgb}{0.3,0,0.5}
\definecolor{dkred}{rgb}{0.5,0,0}

\usepackage{etoolbox}
\usepackage{mdframed}

\usepackage{color}
\definecolor{light-gray}{gray}{0.95}

\usepackage[nottoc,notlof,notlot]{tocbibind} 
\usepackage{hyperref}
\hypersetup{
    colorlinks,
    citecolor=blue,
    filecolor=blue,
    linkcolor=blue,
    urlcolor=blue
}
\usepackage{wrapfig}
\usepackage{algorithmic}
\usepackage[]{algorithm2e}
\usepackage{breakurl}

\newcommand{\hbra}{
  \hbox to \linewidth{\vrule width0.3mm height 1.8mm depth-0.3mm
    \leaders\hrule height1.8mm depth-1.5mm\hfill
    \vrule width0.3mm height 1.8mm depth-0.3mm}}
\newcommand{\hket}{
  \hbox to \linewidth{\vrule width0.3mm height1.5mm
    \leaders\hrule height0.3mm\hfill
    \vrule width0.3mm height1.5mm}}

\title{Randomness of Spritz via DieHarder testing}
\author{Răzvan Ro\c{s}ie 
\\
\\
\\ }

\date{} 

\begin{document}
\maketitle

\begin{abstract}
{\bf RC4} is a stream cipher included in the TLS protocol, and widely used for encrypting
network traffic during the last decades. {\bf Spritz} is a possible candidate for replacing RC4.
Spritz is based on a sponge construction and preserves the byte-oriented behaviour
existing in RC4, but introduces an interface that provides  encryption,
hashing or MAC-generation functionalities.  

We present here the results obtained after applying several statistical tests
on the keystreams generated by Spritz when used in the cipher mode. 
Our methodology makes use of 1024 
keystreams of $2^{25}$ bits. The algorithm
was tested against the DieHarder
test suite. None of the tests failed. Few tests
produced weak results that were corrected when the 
number of samples increased.
\newline


\end{abstract}

\section{Introduction}
\label{Introduction}

Symmetric-key cryptography contains two fundamental classes of cryptographic algorithms: 
{\bf block} ciphers and {\bf stream} ciphers.
During the past decade, symmetric block ciphers were intensively studied
through the AES-election contest organized by NIST. The winning algorithm
is considered secure and has been successfully used in practice. 
On the other hand, less attention was dedicated to the analysis of stream ciphers.
The last contest organized by NIST targeted a different set of cryptographic
primitives: hash functions.

Stream ciphers are used in encrypting mobile communications as well as network 
traffic. Many existing algorithms follow a design of simplicity to favour speed.
The only information-theoretically secure cipher, the One Time Pad cipher (OTP),
serves as an inspirational model for most stream ciphers:
the ciphertext is the result of XOR-ing the plain-text with a random key.
Since the OTP is impractical due to the necessary length of the key, 
various approaches are used by stream ciphers to extend a short key
to possibly infinitely long keystreams that are applied to the plain-text.

RC4 is one of the stream ciphers included into the SSL and TLS cipher suites 
for encrypting Internet traffic, and according to many sources \cite{Sch, Popov}, RC4 is one
of the most popular choices.
Designed in 1987, it was subject to many theoretical and practical attacks; 
a new proposal to replace RC4 was made by Rivest and Schuldt in 2014 - {\bf Spritz} \cite{Spritz}.
The proposed stream cipher is based on a sponge construction, but inherits from the byte-oriented
design of RC4.

\subsection{Motivation}
Few analysis of the algorithm exist \cite{Zoltak}. Here we investigate a preliminary 
aspect of {\bf Spritz}, the randomness of the keystreams that are generated,
independently by the tests that were used by the architects. 
In general, a suspect result obtained when testing the randomness of the output generated
by a stream cipher or a hash function does not imply the existence of an easily exploitable
structural weakness. However, this may jeopardize the chances of the algorithm
to be adopted as a standard.


\section{Specification of Spritz}
\label{Specification}

As described by its authors in \cite{Spritz}, {\bf Spritz} is 
constructed having in mind the benefits offered by sponge functions \cite{Sponge}.
The result is an algorithm that can be used both as a  
message authentication code, random bit generator or a hash function.
However, the primary intention is to use the algorithm and relevant interfaces as a stream cipher.

Spritz follows the standard approach, where the key is extended and added
to the plain-text in order to obtain the cipher-text. 
A $state$ of the algorithm consists of the values of six registers $i,j,k,w,z,a$ and the permutation $S$
of $\{ 0,1,.., N-1\}$.
The default size of $S$ is $N=256$.
As for sponge constructs, the output is generated by a {\bf Squeeze} procedure, which repeatedly
calls a pseudo-random function - {\bf Drip} - that updates the state (changes the values of $i,j,k$ registers and
acts on the permutation $S$) in order to output a byte.
When used as a stream cipher, the key is firstly absorbed into the state through blocks of 'nibbles' (half bytes)
and executing a {\bf Shuffle} procedure after each absorption. The functionality behind this procedure
is aimed to randomize the state. {\bf Shuffle} calls specific procedures that act on the state of the algorithm.

The byte-oriented design the algorithm has the disadvantage of generating the output slower
than other word-oriented algorithms, as remarked in \cite{Spritz}. 
Also, Spritz used in hash mode is slower than
the recently elected SHA-3. \footnote{The width of the permutation in 
SHA-3's winning proposal is proportional to the size of the word, thus easily adaptable for different architectures.}
\newpage

The pseudocode of the algorithm \cite{Spritz}
 and the interface for using it as a stream cipher are provided below:


$$\hbra$$
InitializeState($N$)
\begin{algorithmic}[1]
\STATE $i = j  = k =z = a = 0$
\STATE $w = 1$
\FOR {$v= 0$  {\bf to} $N -1$} 
        \STATE $S[v] = v$
\ENDFOR 
\end{algorithmic}
Shuffle()
\begin{algorithmic}[1]
\STATE Whip($2N$)
\STATE Crush()
\STATE Whip($2N$)
\STATE Crush()
\STATE Whip($2N$)
\STATE $a=0$
\end{algorithmic}
Whip($r$)
\begin{algorithmic}[1]
\FOR {$v= 0$  {\bf to} $r -1$} 
        \STATE Update()
\ENDFOR 
\STATE w = w+1
\STATE {\bf do} w = w+1
\STATE {\bf until} $GCD(w,N)=1$
\end{algorithmic}
Crush()
\begin{algorithmic}[1]
\FOR {$v= 0$  {\bf to} $\lfloor N/2 \rfloor -1$} 
	\IF {$S[v] > S[N-1-v]$}
         		\STATE Swap($S[v], S[N-1-v]$)
	\ENDIF
\ENDFOR 
\end{algorithmic}
Squeeze($r$)
\begin{algorithmic}[1]
\IF {$a>0$}
 	\STATE Shuffle()
\ENDIF
\STATE $P$ = Array.New($r$)
\FOR {$v= 0$  {\bf to} $r -1$} 
         	\STATE $P[v]$ = Drip()
\ENDFOR 
\STATE {\bf return} $P$
\end{algorithmic}
Drip()
\begin{algorithmic}[1]
\IF {$a>0$}
 	\STATE Shuffle()
\ENDIF
\STATE Update()
\STATE {\bf return} Output()
\end{algorithmic}
Update()
\begin{algorithmic}[1]
\STATE $i = i+ w$
\STATE $j=k+S[j+S[i]]$
\STATE $k=i+k+S[j]$
\STATE Swap($S[i], S[j]$)
\end{algorithmic}
Output()
\begin{algorithmic}[1]
\STATE $z=S[j+S[i+S[z+k]]]$
\STATE {\bf return} $z$ 
\end{algorithmic}
$$\hket$$
\newpage

$$\hbra$$
Absorb($I$)
\begin{algorithmic}[1]
\FOR {$v= 0$  {\bf to} $ I.length -1$} 
        \STATE AbsorbByte($I[v]$)
\ENDFOR 
\end{algorithmic}
AbsorbByte($b$)
\begin{algorithmic}[1]
\STATE AbsorbNibble(LOW($b$))
\STATE AbsorbNibble(HIGH($b$))
\end{algorithmic}
AbsorbNibble($x$)
\begin{algorithmic}[1]
\IF {$a = \lfloor N/2 \rfloor $}
         \STATE Shuffle()
\ENDIF
\STATE Swap($S[a], S[\lfloor N/2 \rfloor +x]$)
\STATE $a = a+1$
\end{algorithmic}
AbsorbStop()
\begin{algorithmic}[1]
\IF {$a = \lfloor N/2 \rfloor $}
         \STATE Shuffle()
\ENDIF
\STATE $a = a+1$
\end{algorithmic}
$$\hket$$
\textrm{}
\textrm{}

The following functions use the sponge to
provide hashing, encryption or decryption functionalities. 
$$\hbra$$
Encrypt($K,M$)
\begin{algorithmic}[1]
\STATE KeySetup(K)
\STATE $C = M + $ Squeeze($M.length$)
\STATE {\bf return} $C$ 
\end{algorithmic}
Decrypt($K,C$)
\begin{algorithmic}[1]
\STATE KeySetup(K)
\STATE $M = C - $ Squeeze($C.length$)
\STATE {\bf return} $M$ 
\end{algorithmic}
Hash($M, r$)
\begin{algorithmic}[1]
\STATE InitializeState()
\STATE Absorb($M$)
\STATE AbsorbStop()
\STATE Absorb($r$)
\STATE {\bf return} Squeeze($r$) 
\end{algorithmic}
EncryptWithIV($K,IV,M$)
\begin{algorithmic}[1]
\STATE KeySetup($K$)
\STATE AbsorbStop()
\STATE Absorb($IV$)
\STATE $C = M + $ Squeeze($M.length$)
\STATE {\bf return} $C$ 
\end{algorithmic}
KeySetup($K$)
\begin{algorithmic}[1]
\STATE InitializeState()
\STATE Absorb($K$)
\end{algorithmic}
$$\hket$$
\newpage

\section{Statistical testing of output generated by Spritz}
According to \cite{Spritz, Zoltak}, the algorithm went through extensive
statistical testing. The design rationale  is detailed in the specification 
(including the possible candidates and the choices made for $Update$ and $Output$).
However, the results of the statistical randomness tests applied 
on the keystreams generated by Spritz were not included in the paper.
The randomness of the output is a key feature every stream cipher must have; thus
we expect Spritz to pass all tests, and confirm the claims
made by the authors.
In this section, we present the methodology and the results
obtained by inspecting the keystreams using the 
{\bf DieHarder} suite \cite{DieHarder}.

\subsection{Methodology}
Spritz was used in the stream-cipher mode.
In order to obtain the output for feeding DieHarder, we initially generated the keystreams needed by the Encrypt 
procedure (without initial values).
We used randomly generated keys of 32 bytes length. 
The size of the permutation was 256 bytes (default value).
The data generated consisted of 1024 keystreams of $2^{25}$ bits.
The test environment used was a Linux AMI micro-instance, available from  
Amazon Web Services. 

We motivate our choice for using the DieHarder test suite, due to the fact that it includes
most of the tests existing in NIST's Statistical Test Suite \cite{STS} and the
DieHard battery of tests \cite{DieHard}. Also, the test suite allows to keep testing 
the data until a result is reached with high confidence.

\subsection{Tests used in the statistical analysis}
DieHarder is a comprehensive tool in terms of the statistical tests
that are incorporated.  The recent versions fully include the NIST suite consisting 
of  tests for measuring the frequency, block frequency, entropy, runs, matrix rank, longest run, overlapping or non-
overlapping template matchings,  linear complexity, serial cumulative sums,
random excursions and variants. We do not provide the description of these tests.
However we provide the description of particular tests  from the DieHarder randomness battery that revealed 
a weak behaviour for smaller number of samples: the Monobit, Serial,
RGB Bit Distibution and RGB Permutation tests:
\newline

{\bf The Monobit Test}
\begin{quote}
``Counts the 1 bits in a long string of random uints.  Compares to expected number, generates a p-value directly from 
erfc(). Very effective at revealing overtly weak generators; Not so good at determining where stronger ones eventually 
fail." \cite{DieHarder}
\end{quote}

\newpage

\begin{table}[ht]
\caption{Results for 1024 keystreams of length $2^{25}$.} 
\centering 
\begin{tabular}{c c c c c} 
\hline\hline 
Test Name & tuple & psample& p-value & Result\\ [0.5ex] 
\hline\hline 
diehard\_birthdays & 0 & 100 & 0.63054096 & Passed\\ [0.5ex] 
diehard\_operm5 & 0 & 100& 0.1282933 & Passed\\ [0.5ex] 
diehard\_rank_32x32 & 0 &100& 0.41084472 & Passed\\ [0.5ex] 
diehard\_rank_6x8 & 0 & 100&  0.29555723 & Passed\\ [0.5ex] 
diehard\_bitstream & 0 & 100& 0.22681182 & Passed\\ [0.5ex] 
diehard\_opso & 0 & 100 & 0.95338252 & Passed\\ [0.5ex] 
diehard\_oqso & 0 & 100&0.08739302 & Passed\\ [0.5ex] 
diehard\_dna & 0 & 100& 0.88528270 & Passed\\ [0.5ex] 
diehard\_count\_1s\_str & 0&100 & 0.17277707 & Passed\\ [0.5ex] 
diehard\_count\_1s\_byt & 0&100 & 0.68141261 & Passed\\ [0.5ex] 
diehard\_parking\_lot & 0 & 100&0.08478625 & Passed\\ [0.5ex] 
diehard\_2d\_sphere & 2&100 & 0.81343673 & Passed\\ [0.5ex] 
diehard\_3d\_sphere & 3&100 & 0.84124724 & Passed\\ [0.5ex] 
diehard\_squeeze & 0 &100& 0.59072752 & Passed\\ [0.5ex] 
diehard\_sums & 0 &100& 0.24236920 & Passed\\ [0.5ex] 
sts\_monobit & 0 &100& 0.99792691 & Weak\\ [0.5ex]
sts\_monobit & 0 &200& 0.76653420 & Passed\\ [0.5ex] 
sts\_runs & 0 & 100&0.97730946 & Passed\\ [0.5ex] 
sts\_serial & 1&100 & 0.99792691 & Weak\\ [0.5ex] 
sts\_serial & 2&100 & 0.89538406 & Passed\\ [0.5ex] 
sts\_serial & 3&100 & 0.62255692 & Passed\\ [0.5ex] 
sts\_serial & 3 &100& 0.97791641 & Passed\\ [0.5ex]
sts\_serial & 4&100 & 0.98491218 & Passed\\ [0.5ex] 
sts\_serial & 4&100& 0.72518227 & Passed\\ [0.5ex] 
rgb\_bitdist & 1&100 & 0.97541932 & Passed\\ [0.5ex]
rgb\_bitdist & 2&100& 0.99908738 & Weak\\ [0.5ex]
rgb\_bitdist & 2&200& 0.99908738 & Passed\\ [0.5ex]
rgb\_bitdist & 3&100 & 0.60061288 & Passed\\ [0.5ex]
rgb\_bitdist & 4&100& 0.95070961 & Passed\\ [0.5ex]
rgb\_bitdist & 5&100& 0.970557854& Passed\\ [0.5ex]
rgb\_permutations  & 2&100 & 0.99583035& Weak\\ [0.5ex]
rgb\_permutations  & 2&200 & 0.49210341& Passed\\ [0.5ex]
rgb\_permutations  & 3&100 & 0.34685706& Passed\\ [0.5ex]
rgb\_permutations  & 4&100 & 0.10607070& Passed\\ [0.5ex]
rgb\_permutations  & 5&100 & 0.47826848& Passed\\ [0.5ex]
rgb\_lagged\_sum  & 0&100 & 0.48016287 & Passed\\ [0.5ex]
rgb\_lagged\_sum  & 1 &100& 0.05039327 & Passed\\ [0.5ex]
rgb\_lagged\_sum  & 2 &100& 0.10018457& Passed\\ [0.5ex]

\hline 
\end{tabular}

\label{table:nonlin} 
\end{table}
\newpage

{\bf The Serial Test}
\begin{quote}
``Accumulates the frequencies of overlapping n-tuples of bits drawn from a source of random integers. The expected 
distribution of n-bit patterns is multinomial with $p$ = $2^{-n}$ e.g. the four 2-bit patterns 00 01 10 11 should occur with 
equal probability." \cite{DieHarder}
\end{quote}

{\bf The RGB Bit Distribution Test}
\begin{quote}
``Accumulates the frequencies of all n-tuples of bits in a list of random integers and compares the distribution thus 
generated with the theoretical (binomial) histogram, forming chisq and the associated p-value. In this test n-tuples are 
selected without overlap (e.g. 01|10|10|01|11|00|01|10) so the samples are independent." \cite{DieHarder}
\end{quote}

{\bf The RGB Permutation Test}
\begin{quote}
``This is a non-overlapping test that simply counts order permutations of random numbers, pulled out n at a time. 
There are  n! permutations and all are equally likely. 
The samples are independent, so one can do a simple chisq test on the count 
vector with n! - 1 degrees of freedom." \cite{DieHarder}
\end{quote}

\subsection{Results}
The results obtained after running the DieHarder tests confirm the
claims of the specification of the algorithm. The $p$-values obtained from the
tests were all greater than 0.99, except for the Monobit, Serial, RGB Bit Distribution
and RGB Permutation tests. When the number of $p$-$samples$
was increased to 200, the previous tests passed.

Also, a close to 0.01 value was identified for several other tests. 
To improve the accuracy of the results
a larger amount of data may be needed and the number of $p$-$samples$
may be increased as well.

\section{Conclusion}
This report investigates the randomness of the output generated by  Spritz through statistical means.
We applied  the DieHarder test battery over a set of keystreams produced by the algorithm.
The results do not indicate any failure of a randomness statistical test. 
Suspect $p$-$values$ are observed for few tests. This behaviour
is completely eliminated when DieHarder  increases the number of samples and repeats the tests -
a fact confirms the initial expectations - no non-random behaviour
was observed in Spritz.


\subsection{Future Work}
Spritz is an interesting framework, given the large range of possible applications.
As mentioned in the proposal, additional investigations
are needed for Spritz operating as a hash function, deterministic random bit generator
or message authentication code provider. A future investigation can 
analyze the algorithm statistically and structurally when used in these modes. 

Other improvements can be summarized in testing larger input files or using different statistical tests,
that are not available in DieHarder. Many of these tests may be performed in a parallel
way. An existing work \cite{Kam} provides a testing methodology for hash functions.
In future, this can be used to investigate the randomness of Spritz when used
in the hash mode.

\newpage

\end{document}